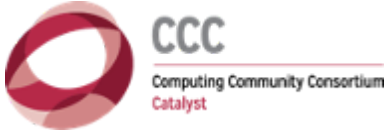

# An Agenda for Disinformation Research

*A Computing Community Consortium (CCC) Quadrennial Paper*

*Nadya Bliss (Arizona State University), Elizabeth Bradley (University of Colorado, Boulder), Joshua Garland (Santa Fe Institute), Filippo Menczer (Indiana University), Scott W. Ruston (Arizona State University), Kate Starbird (University of Washington), and Chris Wiggins (Columbia University)*

In the 21$^{st}$ Century information environment, adversarial actors use disinformation to manipulate public opinion. The distribution of false, misleading, or inaccurate information with the intent to deceive is an existential threat to the United States—distortion of information erodes trust in the socio-political institutions that are the fundamental fabric of democracy: legitimate news sources, scientists, experts, and even fellow citizens. As a result, it becomes difficult for society to come together within a shared reality; the common ground needed to function effectively as an economy and a nation.

Computing and communication technologies have facilitated the exchange of information at unprecedented speeds and scales. This has had countless benefits to society and the economy, but it has also played a fundamental role in the rising volume, variety, and velocity of disinformation. Technological advances have created new opportunities for manipulation, influence, and deceit. They have effectively lowered the barriers to reaching large audiences, diminishing the role of traditional mass media along with the editorial oversight they provided.

The digitization of information exchange, however, also makes the practices of disinformation *detectable*, the networks of influence *discernable*, and suspicious content *characterizable*. New tools and approaches must be developed to leverage these affordances to understand and address this growing challenge. Tools must be developed for security agencies, educators, journalists, civil society organizations, and citizens at large to make sense of, and counter, information pollution. These solutions must incorporate better understandings and models of the "demand side" of the disinformation ecosystem—the consumers of the content—as much as the detection, attribution and characterization efforts support recognition and interdiction on the "supply side," where it originates. Development of such tools and approaches will require collaboration of computer and computational scientists with cognitive and social scientists to better understand this ecosystem and model vulnerabilities in a comprehensive way. As a research topic, the disinformation landscape is a socio-technical ecosystem; research approaches need to meet the new challenges of such a landscape, including adversarial actors and platform companies whose product decisions shape the nature of the threat and its diffusion. Critically, all disinformation solutions must respect ethical principles that balance the privacy and autonomy of individuals online with the societal benefits of understanding and mitigating the threat.

**Recommendations**

To meet the disinformation challenge, we need investments in fundamental research as well as interventions that knit together the social and the technical, that solve the measurement/impact quantification task, and that effectively balance ethical trade-offs around appropriate uses of personal data.

We identify *six strategic targets of support* that will move us toward these ambitious goals: detection of disinformation at scale, measurement of impact, data infrastructure, educational interventions, workforce training, and new ethical guidelines.

1. **Detection of disinformation at scale:** Disinformation and manipulation are adversarial challenges, so both the types of abuse and the methods for detecting them will continue to evolve in the foreseeable future. Inauthentic actors (social bots, trolls, cyborgs, etc.), manipulated and synthetic media artifacts (video, speech, text), coordinated information operations (foreign and domestic), astroturf campaigns, false and/or misleading claims, conspiracy theories, junk science, and fake news sources are all examples of phenomena that have been exacerbated by technology in recent years. Research efforts must be directed not only toward detection, but also the formidable challenges posed by provenance, attribution, integrity, and verification. Advances in AI, from Generative Adversarial Networks (GAN) to create deepfakes to Generative Pre-trained Transformers (GPT) to synthesize text, are becoming ripe for empowering bad actors faster than for developing countermeasures. Serious effort is needed to reverse these trends. Advances in knowledge graphs, the semantic web, machine learning, networking, and data science need to be targeted toward both the detection and propagation of online manipulation at scale.
2. **Measurement of impact:** Research is needed to measure the impact of disinformation in different cultural and geographic contexts, over long periods of time, and taking into account second-order effects on social norms, ideologies, epistemologies and sociotechnical structures (like algorithms and social networks) that mediate these impacts. Precise, reliable, and validated measurement of the "effect" or "impact" of disinformation on communities requires formal statistical causal inference on human belief dynamics. Such calculations are currently a challenge due to the range of independent variables and difficulties in quantifying them. Solutions to this challenge will require advances in the identification and extraction of complex cognitive/rhetorical structures (e.g., metaphors, narratives, frames) and in the development of enduring laboratory proxies of human community engagement—i.e., some form of experimental "sandbox" that is deeply representative of the ecosystem and freely available to researchers.
3. **Data infrastructure:** We need a common research infrastructure to access data from technology platforms under ethical guidelines that protect user privacy and transparent administrative rules that protect intellectual property. Different platforms should provide vetted researchers with comparable, open APIs to enable cross-platform analyses of disinformation. Such a collaboration with technology platforms cannot be left to individual researchers without harming reproducibility and replicability. Incentives must be provided for platforms and researchers to collaborate across institutions and fields while respecting the motivations of the different sectors. In addition, we need to develop protocols and large-scale infrastructures that allow citizens to contribute data for research in a secure and privacy-preserving manner.

4. **New ethical guidelines:** Any experimental research aimed at detecting and characterizing disinformation requires gathering of data about real individuals and their communications. In this, the privacy of those individuals must be respected and potential harms must be anticipated. Furthermore, direct measurement of the causal effects of disinformation on private platforms requires intervention: transparency, fairness, and minimization of harms must be ensured. The ethical standards and practices of Human Subjects Research (HSR) are currently guided by the 1978 Belmont Report. Its principles are interpreted via individual institutional review boards, primarily at research universities, often with differing interpretations. For example, some review boards consider the collection of any digital image with human faces subject to HSR protections regardless of the research focus and method; other ethical boards do not consider the collection of such images as requiring HSR protocols. Whether public Twitter content is subject to privacy protections is also inconsistent. Such discrepancies hinder collaboration between university, industry, and government teams, and can unduly restrict research from achieving valuable results while not appreciably enhancing protections. Policymakers have an opportunity to commission a report of similar impact to the original Belmont report, updating how its ethical principles should be interpreted in the very different context of today's disinformation ecosystem.

5. **Educational interventions:** Detection efforts are important interventions on the supply side of disinformation. On the demand side, we need to better prepare our citizens for dealing with the modern, computationally accelerated and algorithmically driven information environment. Not only do few people outside of social media platforms understand how prioritization algorithms work; few people even understand that algorithms are at play, and that their purpose is not to inform but to commodify and influence the user. This has caused considerable damage throughout our society, resulting in large-scale polarization and conspiracy theories, such as the anti-vaccine movement. In the face of these pressures, we need broad educational initiatives that raise the level of fundamental knowledge about the information environment to enable citizens to adequately function in civil society. These interventions will require research at the intersection between psychology, sociology, philosophy, and the computer and information sciences. Additionally, we need to develop tools that journalists, scientists, and educators can leverage to underpin credible information.

6. **Workforce training:** Much of the technology that is being blamed today for disinformation and manipulation was developed with benign intent and initially brought significant benefits. Its negative repercussions and weaponization were not foreseen by the technologists who developed it. This sheds light on the question of how best to train the next generation of computing professionals in such a way that their processes and aspirations align not only with technical excellence, but with a practicable mindset and toolset for applied ethics. Policymakers and research funders sit at a unique point in terms of both perspective and impact in this conversation. Focused and thoughtful programs should be developed and deployed, along with technical training, to ensure that future generations of computing professionals not only understand the importance of applied ethics in their work, but more importantly possess a shared, useful vocabulary and conceptual toolkit for recognizing and facing future ethical challenges. At present we can only speculate about these challenges, as technology continues to change and grow in its power and influence over our realities.

This important and ambitious agenda will require a blend of humanities, social science, education, journalism, and computer and science, with comprehensive support and participation from a broad range of organizations and institutions, including:

- The initiation of dedicated interdisciplinary research programs at the National Science Foundation;
- The creation of public-private partnerships to support accessible research infrastructure;
- The fostering of cross-agency collaboration, especially between the National Science Foundation, the Department of Defense R&D, the Department of Homeland Security Science & Technology Directorate, and the Intelligence Community's Science & Technology efforts, to support the transition of promising research outcomes and secure the integrity of the information ecosystem;
- The formation of cross-agency/cross sector partnerships—with engagement from the media industry, among others—to support education and workforce training initiatives; and perhaps most crucial:
- The active, transparent, and good faith participation of the platform companies, whose algorithms and product decisions shape the spread and amplification of disinformation online.

As the scale and breadth of online information content and creation continue to grow, and as more and more of our society moves online, this interdisciplinary agenda for research, education, and infrastructure will be vital in preserving our democratic society and mitigating the unintended consequences of our technological advancement. Focused and sustained investments, along with programmatic implementation strategies, as outlined above, present a clear path forward towards a trustworthy information ecosystem.

*This white paper is part of a series of papers compiled every four years by the CCC Council and members of the computing research community to inform policymakers, community members and the public on important research opportunities in areas of national priority. The topics chosen represent areas of pressing national need spanning various subdisciplines of the computing research field. The white papers attempt to portray a comprehensive picture of the computing research field detailing potential research directions, challenges and recommendations.*

*This material is based upon work supported by the National Science Foundation under Grant No. 1734706. Any opinions, findings, and conclusions or recommendations expressed in this material are those of the authors and do not necessarily reflect the views of the National Science Foundation.*